\documentclass{article}

\usepackage{arxiv}
\usepackage[utf8]{inputenc} % allow utf-8 input
\usepackage[T1]{fontenc}    % use 8-bit T1 fonts
\usepackage{hyperref}       % hyperlinks
\usepackage{url}            % simple URL typesetting
\usepackage{booktabs}       % pro]pfessional-quality tables
\usepackage{amsfonts}       % blackboard math symbols
\usepackage{nicefrac}       % compact symbols for 1/2, etc.
\usepackage{microtype}      % microtypography
\usepackage{lipsum}
\usepackage{afterpage}
\usepackage{graphicx}
\usepackage{color}
\usepackage{amsmath}
\usepackage{cancel}
\usepackage{caption}
\usepackage{amssymb}
\usepackage{cite}

\title{Dark matter constraints and the neutralino sector of the scNMSSM}

\author{
   Elham ~Aldufeery, and Maien Y.~Binjonaid \footnote{maien@ksu.edu.sa}\\
  Department of Physics and Astronomy\\
  King Saud University\\
  Riyadh, Saudi Arabia \\
}

\begin{document}
\maketitle

\def\beq{\begin{equation}}   \def\eeq{\end{equation}}
\def\ba{\begin{array}}       \def\ea{\end{array}}
\def\bea{\begin{eqnarray}}   \def\eea{\end{eqnarray}}
\def\nn{\nonumber}
\def\nl{\newline}
\def\k{\kappa}
\def\l{\lambda}
\def\b{\beta}
\def\t{\theta}
\def\noi{\noindent}

\begin{abstract}
The neutralino sector of the semi-constrained next-to-minimal supersymmetric standard model is explored under recent experimental constraints, with special attention to dark matter (DM) limits. The effects of the upper and lower bounds of dark matter relic density and recent direct detection constraints on spin-independent and -dependent cross-sections are thoroughly analyzed. Particularly, we show which regions of the parameter space are ruled out due to the different dark matter constraints and the corresponding model-specific parameters: $\lambda, \kappa, A_{\lambda}$, and $A_{\kappa}$. We analyze all annihilation and co-annihilation processes (with heavier neutralinos and charginos) that contribute to the dark matter relic density. The mass components of the dark matter candidate, the lightest neutralino $\tilde{\chi}_1^0$, are studied, and the decays of heavy neutralinos and charginos, especially $\tilde{\chi}_2^0$ and $\tilde{\chi}_1^+$, into the lightest neutralino are examined. We impose semi-universal boundary conditions at the Grand Unified Theory scale, and require a moderate range of $\tan{\beta} \lesssim 10$. We find that the allowed parameter space is associated with a heavy mass spectrum in general and that the lightest neutralino is mostly Higgsino with a mass range that resides mostly between 1000 and 1500 GeV. However, smaller mass values can be achieved if the DM candidate is bino-like or singlino-like. 


\end{abstract}

\section{Introduction} \label{intro}
The nature of dark matter (DM) continues to be a mystery, even though there is an abundance of evidence of its existence \cite{Bertone:2016nfn, Corbelli:1999af, Faber:1976sn, Allen:2011zs, Refregier:2003ct, Ade:2015xua}. A well-motivated candidate for DM is the neutralino lightest supersymmetric particle (LSP) in supersymmetric (SUSY)  
models \cite{Zyla:2020zbs}. However, such models are being challenged by recent null results from experiments, including DM searches \cite{ATLAS:2019oho,ATLAS:2020jhr,Lorenz:2019pxf,ATLAS:2020ckz,Aad:2019vvi, Sirunyan:2020zzv}. DM searches can be categorized into collider, direct detection, and indirect detection searches. Collider searchers rely on detecting missing energy events that could arise from decays into DM particles; indirect detection depends on observations of astronomical events, such as gamma ray lines; and direct detection searches rely on observing possible elastic scattering events between DM particles and atomic nuclei (a comprehensive review was published \cite{Bertone:2004pz}). One of the most-considered low-scale SUSY models is the next-to-minimal supersymmetric standard model (NMSSM; reviewed in \cite{Ellwanger:2009dp, Maniatis:2009re}), which opens new possibilities as its neutralino sector is richer than that in the MSSM, which is attributed to the inclusion of a Standard Model (SM)
singlet superfield, which is absent in the MSSM. The inclusion of such a singlet is motivated by explaining the origin of the $\mu$ term in the MSSM and increasing the predicted SM Higgs mass at the tree level, especially for the case where the singlet--doublet coupling $\lambda$ is large and $\tan_{\beta}$ is small (see, e.g., \cite{Cao:2012fz, Binjonaid:2014oga} and the references therein). 

The issue of DM in the NMSSM is the focus of the attention of researchers in the field \cite{Abel:1992ts, Stephan:1997ds, Stephan:1997rv, Hugonie:2007vd, Belanger:2005kh, Belanger:2008nt, Kraml:2008zr, Yang:2010zzd, Cao:2011re, Ellwanger:2014dfa,Cao:2013mqa, Jeong:2014xaa, Wang:2016lvj, Badziak:2015exr, Cao:2018rix, Wang:2019biy, Li:2020ufo, Wang:2018vxp, Wang:2020dtb}. The case of low-mass neutralino LSP has been investigated with attention paid to the interesting cases of large singlino--higgsino mixtures or singlino-like DM \cite{Kraml:2008zr, Ellwanger:2014hia, Potter:2015wsa, Ellwanger:2018zxt, Ellwanger:2016sur, Guchait:2020wqn, Barman:2020zpz}. Additionally, the implications of direct and indirect detection of the NMSSM and its extensions have been detailed \cite{Cerdeno:2004xw, Belanger:2008nt, Das:2010ww, AlbornozVasquez:2012px, Beskidt:2017xsd, Cao:2019aam, Cao:2019qng, Ferrer:2006hy, Han:2014nba}. The existence of blind spots, leading to a reduction in spin-independent direct detection cross-sections has been studied \cite{Badziak:2015exr, Baum:2017enm}. Recently, Ref.~\cite{Barman:2020vzm} studied the NMSSM and DM, where low-scale input parameters were used, focusing on future searches. Additionally, certain scenarios of DM in the semi-constrained NMSSM (scNMSSM), which is the model we are considering, were recently investigated in \cite{Wang:2018vxp, Wang:2020tap}. However, most of these works do not include the most recent constraints from DM searches, especially those from direct detection. Particularly, Ref.~\cite{Wang:2018vxp} did not include constraints from DarkSide-50 \cite{Agnes:2018ves}, CRESST-II \cite{Angloher:2015ewa}, XENON1T \cite{Aprile:2019dbj} and PICO60 \cite{Amole:2019fdf}, which are included in Ref.~\cite{Wang:2020tap}. However, in the latter reference, only cases were $m_{\tilde{\chi}^0_1} < m_{h_{\text{SM}}}/2 $ are considered. Both papers focus on a limited portion of the parameter space of the scNMSSM, which is motivated by satisfying $(g-2)_{\mu}$, and only the upper limit on the dark matter relic density is taken into account. Moreover, the Higgs sector of both papers is associated with $h_2$ being the SM-like Higgs boson, which is different from the parameter space to be considered in this paper. Some aspects of the Higgs sector of the parameter space we are considering has been analyzed in a recent paper \cite{Binjonaid:2020mpla} where only the upper limit on DM was imposed.

From a phenomenology point of view, it is interesting to understand in detail how current limits affect the NMSSM with the well-motivated Grand Unified Theory (GUT) boundary conditions in terms of the affected regions of the parameters and, more physically, which channels contribute to DM relic density, which is the branching ratio of chargino/neutralino to the LSP. Considering the above, we explored 
the parameter space of the NMSSM under the semi-constrained boundary conditions on the GUT scale exploiting state-of-the-art tools incorporating the latest DM constraints, as a major motivation for SUSY models is to provide gauge coupling unification. Particularly, we analyzed how relic density abundance constraints, as well as recent direct detection limits, affect the model, without a particular focus on specific regions, except for relatively moderate $\tan{\beta} \lesssim 10$, which was motivated by enhancing the tree-level prediction of the SM-like Higgs mass in the model. We present the portions of the parameter space that pass all known constraints (apart from g-2 of the muon), and show the regions that fail the aforementioned DM constraints, providing a deeper insight into the model that has usually been overlooked in the recent literature.

The paper is organized as follows: In Section \ref{sec:neutsec}, we provide a brief overview of the superpotential of the NMSSM and the neutralino mass matrix, and introduce the necessary parameters of the model. Next, Section \ref{sec:scan} specifies the boundary conditions of model and presents our methodology for probing the parameter space of the model, and the relevant non-DM and DM constraints that are considered. In Section \ref{sec:results}, the results are presented and divided into subsections corresponding to the various considered aspects of the neutralino section. Section \ref{sec:discon} discusses the results and compares them to recent literature, and concludes the paper. 

\section{The NMSSM and its Neutralino Sector} \label{sec:neutsec}
The NMSSM is characterized by the following superpotential:
\begin{equation}
 \mathcal{W} = \mathcal{W}_{\text{MSSM}}^{\cancel{\mu}} + \lambda \hat S \hat H_u \cdot \hat H_d + \frac{1}{3} \kappa \hat S^3,
\end{equation}
where $\mathcal{W}_{\text{MSSM}}^{\cancel{\mu}}$ encodes the Yukawa couplings as in the MSSM. The second term replaces the $\mu$ term in the MSSM so that it is dynamically generated when 
the singlet superfield $\hat S$ acquires a vacuum expectation value (VEV). $\lambda$ is the coupling between $\hat S$ and the up and down Higgs superfields $\hat H_u, \hat H_d$. The third term represents the cubic self-coupling of $\hat S$. This $\kappa$ term is included to maintain a $\mathbb{Z}_3$ invariance in the model. The soft-SUSY breaking Lagrangian of the NMSSM is:

\begin{eqnarray}
\nonumber -\mathcal{L}^{\text{\tiny NMSSM}}_{\text{\tiny soft}}&=&\dfrac{1}{2}\left(M_1\tilde{\lambda}_1\tilde{\lambda}_1+M_2\tilde{\lambda}_2^i\tilde{\lambda}_2^i+M_3\tilde{\lambda}_3^a\tilde{\lambda}_3^a+h.c.\right)\\
\nonumber &&+ m_{\tilde{Q}}^2|\tilde{Q}|^2+m_{U}^2|\tilde{U}_{R}|^2+ m_{\tilde{D}}^2|\tilde{D}_{R}|^2+ m_{\tilde{L}}^2|\tilde{L}|^2+m_{\tilde{E}}^2|\tilde{E}_{R}|^2\\
\nonumber&&+m^2_{H_u}\left|H_u\right|^2+ m^2_{H_d}\left|H_d\right|^2+m^2_{S}\left|S\right|^2\\
\nonumber&&+ \bigg(h_uA_u\tilde{U}_{R}^c\tilde{Q}\cdot H_u -h_dA_d\tilde{D}_{R}^c\tilde{Q}\cdot H_d\\
&&-h_eA_e\tilde{E}_{R}^c\tilde{L}\cdot H_d+\lambda A_{\lambda}\ S H_u\cdot H_d+\dfrac{1}{3}\kappa A_{\kappa} S^3 +h.c.\bigg).\label{soft1}
\label{soft}
\end{eqnarray}

The first line of Eq. (\ref{soft}) represents the gaugino mass terms, where $\tilde{\lambda}_1$, $\tilde{\lambda}_2^i$ $(i=1,2,3)$, and $\tilde{\lambda}_3^a$ $(a=1,...,8)$ denote the $U(1)_Y$, $SU(2)$, and $SU(3)$ gauginos, respectively, with corresponding masses $M_1$, $M_2$, and $M_3$. The five mass terms $m_{\tilde{Q}}^2$, $m_{U}^2$, $m_{\tilde{D}}^2$, $m_{\tilde{L}}^2$, and $m_{\tilde{E}}^2$ are $3\times 3$ Hermitian matrices in the generation space and represent the squarks and sleptons mass squared terms, respectively.  
$m^2_{H_u}$ and $m^2_{H_d}$ contain Higgs mass squared terms, and $m^2_{S}$ is the singlet mass term. Here, $\boldsymbol{y_{u,d,e}}$, $\lambda$, and $\kappa$ are dimensionless Yukawa couplings. Finally, the $A$ parameters describe a trilinear scalar interaction and has a dimension of mass. In contrast to the MSSM, the soft-SUSY breaking Lagrangian of the NMSSM contains extra terms associated with trilinear terms $A_{\lambda}$ and $A_{\kappa}$, as well as a soft-SUSY breaking mass term for the singlet. The Higgs sector of the model receives extra contributions from the singlet, and the number of Higgs bosons is increased to seven particles (2 charged, 3 {CP}-even, and 2 CP-odd).  
One of the known advantages of the addition of the singlet superfield is that the physical SM-like Higgs mass receives an additional F-term contribution that is proportional $\lambda$. Specifically,

\begin{equation}
 m_h^2 \leq M_Z^2 \left[\cos^2{(2 \beta)} + \frac{\lambda^2}{g^2} \sin^2{(2 \beta)} \right], 
\end{equation}
which motivates choosing small to moderate values of $\tan{\beta}$ and a sizeable value for $\lambda$ to enlarge the tree-level SM-like Higgs mass, as long as the choice of $\lambda$ does not lead to violating the perturbativity bound. This could occur when $\lambda$ exceeds about $0.7$.

When electroweak symmetry breaking (EWSB) is achieved in the NMSSM, the mass of the Z boson is predicted to be:
\begin{equation}
\dfrac{M_Z^2}{2}=\dfrac{m_{H_d}^2-m_{H_u}^2 \tan^2\beta}{\tan^2\beta-1}-|\mu_{\text{eff}}|^2,
\label{MZ}
\end{equation}
which is usually used to quantify the fine-tuning in the electroweak sector of the model. A widely used fine-tuning measure is the one introduced in \cite{Barbieri:1987fn}, which is defined as:
\begin{equation}
 \Delta_{M_Z}^{\text{GUT}} = \left| \frac{d \ln M_Z^2}{d \ln a^2} \right|,
\end{equation}
where $a$ represents the input parameters of the model on the GUT scale.

The neutralino sector is also affected by the presence of the new NMSSM-specific terms. As opposed to the $4\times4$ mass matrix in the MSSM, the neutralino mass matrix in the NMSSM is $5\times5$, which reads:

\begin{equation}
{\cal M}_0 =
\left( \ba{ccccc}
M_1 & 0 & -\frac{g_1 v_d}{\sqrt{2}} & \frac{g_1 v_u}{\sqrt{2}} & 0 \\
& M_2 & \frac{g_2 v_d}{\sqrt{2}} & -\frac{g_2 v_u}{\sqrt{2}} & 0 \\
& & 0 & -\mu & -\lambda v_u \\
& & & 0 & -\lambda v_d \\
& & & & 2\kappa s
\ea \right),
\end{equation}
where $M_{1,2}$ represent gaugino masses, $\mu = \lambda s$ is the effective $mu$-term, $g_{1,2}$ denotes gauge couplings, $v_{u,d}, s$ indicates the VEVs of the up and down Higgs fields and singlet field. Upon diagonalization, the mass eigenstates are ordered from small to large as $\tilde{\chi}_1^0, \dots, \tilde{\chi}_5^0$, where the first is the lightest neutralino, which is usually the LSP of the model and the dark matter candidate. The lightest neutralino can then be expressed as:
\begin{equation}
\tilde{\chi}_1^0=N_{11}\tilde{B}+N_{12}\tilde{W}^0+ N_{13}\tilde{H}_d^0+N_{14}\tilde{H}_u^0+N_{15}\tilde{S},\\
\end{equation}
where $N_{1j}$ are matrix elements of the diagonalization matrix. They encode the possible mixtures comprising the neutralino mass eigenstates, and $\sum |N_{1j}|^2 =1$. For example, in the bino-like LSP scenario:
\begin{align}
N_{11}&\approx 1,& N_{15}&\approx 0, & N_{13}&\approx \dfrac{m_Z\sin\theta_W}{\mu}\sin\beta,& N_{14}&\approx -\dfrac{m_Z\sin\theta_W}{\mu}\cos\beta,
\end{align}
whereas in the singlino-like LSP scenario:
\begin{align}
N_{11}&\approx 0,& N_{15}&\approx 1, & N_{13}&\approx -\dfrac{\lambda v}{\mu}\cos\beta,& N_{14}&\approx -\dfrac{\lambda v}{\mu}\sin\beta.
\end{align}

Finally, the NMSSM can be specified on the GUT scale by a number of input parameters: a soft-SUSY breaking scalar mass parameter ($m_0$), a soft-SUSY breaking gaugino mass parameter ($m_{1/2}$, a soft-SUSY breaking trilinear parameter $A_0$, the ratio between the VEVs of up and down-Higgs fields $\tan{\beta} = v_u/v_d$, an effective Higgs/higgsino mass parameter $\mu_{\text{eff}} = \lambda s$, a singlet--doublet coupling parameter $\lambda$, and a singlet coupling parameter $\kappa$. It is desirable to understand how NMSSM-specific parameters behave and affect the model's predictions; therefore, the two trilinear soft-SUSY breaking parameters $A_{\lambda}$ and $A_{\kappa}$ are often treated differently from $A_0$.
\section{The parameter space} \label{sec:scan}

NMSSMTools (v5.5.2) was used to generate the mass spectrum of the model. NMSSMTools uses MicrOmega\footnote{The interfacing was performed by the authors of NMSSMTools.}, which in turn calculates the relevant DM observables of the model. All theoretical and experimental constraints were considered apart from the $(g-2)_{\mu}$ anomaly. These constraints include limits on the decays of both the $K$ and $B$ mesons; on the measured SM Higgs mass $\pm 3$ GeV to account for theoretical uncertainties (see \cite{Staub:2015aea} for more details); recent LHC constraints on the measured couplings of the SM Higgs as mentioned, which are imposed in NMSSMTools using the framework presented in \cite{Belanger:2013xza}, and Higgs decays (including exotic decays); and constraints on supersymmetric particles. Theoretical constraints include reaching a global minimum, no negative physical mass values, successful Electroweak symmetry breaking (EWSB) in the Higgs sector.

The package was modified for our purpose, which was to analyze each DM constraint separately after imposing non-DM constraints. This enabled us to understand and report how DM limits affect the considered model. However, as far as DM constraints are concerned, in this paper, we restrict our analysis to constraints on DM relic density and direct detection. 

We explore the parameter space focusing on specific regions where $\tan{\beta} \lesssim 10$ but allow $\lambda$ to vary, as shown in Table \ref{tab:scans}. We specify the inputs of the model on the GUT scale in what is commonly called the semi-constrained next-to-minimal supersymmetric standard model as explained in the previous section. However, since NMSSMTools was used for this study, to find acceptable solutions of the renormalization group equations, this code specifies the boundary conditions at the GUT and low scales simultaneously for practical reasons. In particular, $m_0, m_{1/2}, A_0, A_{\lambda}, A_{\kappa}$ are specified on the GUT scale, whereas $\lambda, \kappa$, and $\tan{\beta}$ are specified on the low scale. $m_{H_u}(GUT)$ and $m_{H_d}(GUT)$ are computed by the code, and $m_{H_u}(GUT)$ and $m_{H_d}(GUT)$ not being set equal to $m_0 (GUT)$ is why this model is usually called semi-constrained and is sometimes called the non-universal Higgs NMSSM (NUH-NMSSM ). More details about how NMSSMTools computes the spectrum and the possible options can be found in the relevant literature \cite{Ellwanger:2006rn,Ellwanger:2005dv,Das:2011dg,Belanger:2014hqa,Domingo:2015qaa}.

\begin{table}[h!]
 \centering
 \begin{tabular}{|c|c|}
 			\hline
 			Parameter & Range \\
 						\hline
 $m_0$ [GUT] & [500--5000] GeV \\
 			\hline
 $m_{1/2}$ [GUT] & [500--5000] GeV \\
 			\hline
 $A_0$, $A_{\lambda}$, $A_{\kappa}$ [GUT] & [--3000-- 3000] GeV \\ 
 			\hline
 $\mu_{\text{eff}}$ [SUSY] & [100--1500] GeV \\
 			\hline
 $\tan{\beta}$ [$M_Z$]& [1--10] \\
 			\hline
 $\lambda$, $\kappa$ [SUSY] & [0.1--0.7] \\
 			\hline
 \end{tabular}
 \caption{Ranges of the input parameters and the input scale. }
 \label{tab:scans}
\end{table}

The ranges of the input parameters are shown in Table \ref{tab:scans}. Notably, the ranges for the trilinear parameters are varied separately, and the same is true for $\lambda$ and $\kappa$. It is worth mentioning that in \cite{Wang:2018vxp, Wang:2020tap}, the maximum allowed values for $m_0, m_{1/2}$ and $\mu_{\text{eff}}$ are 500, 2000 and 200 GeV, respectively. Moreover, only the upper bound on the DM relic density was imposed in both references. This makes our study an extension to such recent works since we consider a much wider ranges for the previously mentioned parameters, and we do not require accounting for $(g-2)_{\mu}$ since it is still considered to be an unconfirmed anomaly. Moreover, in our study, we impose the upper and lower bounds on DM relic density such that the scNMSSM can account for the DM entirely, unlike the previously mentioned references.

Finally, the latest version of NMSSMTools uses MicrOMEGAsv5.0 and includes a number of recent and important constraints on Higgs physics \cite{Aad:2019mbh, Sirunyan:2018koj, Khachatryan:2016vau}, DM limits on spin-independent cross-sections from DarkSide-50 \cite{Agnes:2018ves} and CRESST-II \cite{Angloher:2015ewa}, and limits on spin-dependent cross-sections from XENON1T \cite{Aprile:2018dbl, Aprile:2019dbj} and PICO60 \cite{Amole:2019fdf}.

\section{Numerical Results} \label{sec:results}

\subsection{Mapping DM Constraints into the Model's Parameters}
\begin{figure*}
\centering
 \includegraphics[width=0.85\textwidth]{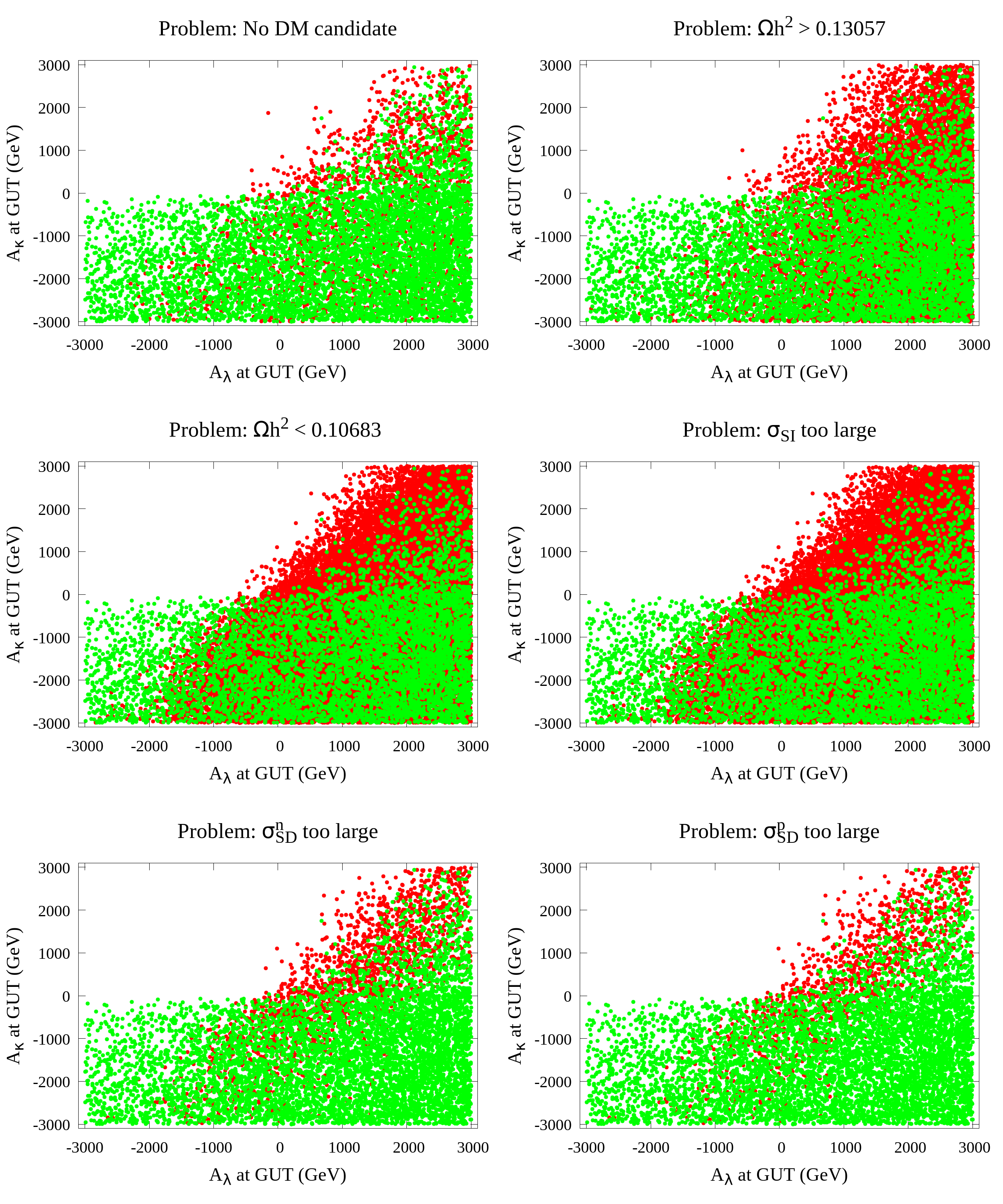}
 \caption{Dark matter (DM) constraints in the $(A_{\lambda}$-$A_{\kappa})$ plane. Red points are ruled out; green points are allowed. }
 \label{fig1}
\end{figure*}

\begin{figure*}
\centering
 \includegraphics[width=0.85\textwidth]{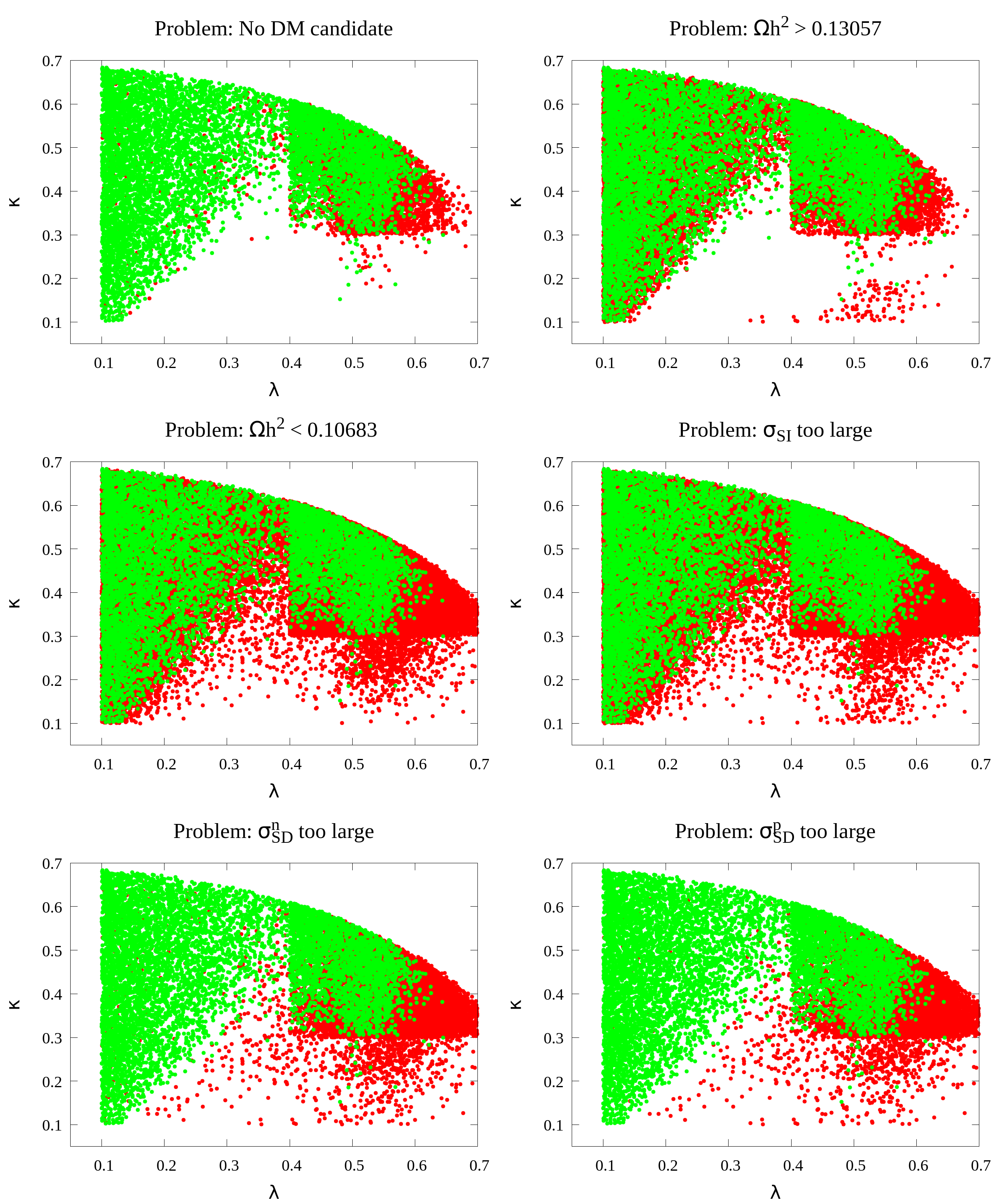}
 \caption{DM constraints in the $\lambda$--$\kappa$ plane. Red points are ruled out; green points are allowed.}
 \label{fig2}
\end{figure*}

\begin{figure*} 
\centering
 \includegraphics[width=0.85\textwidth]{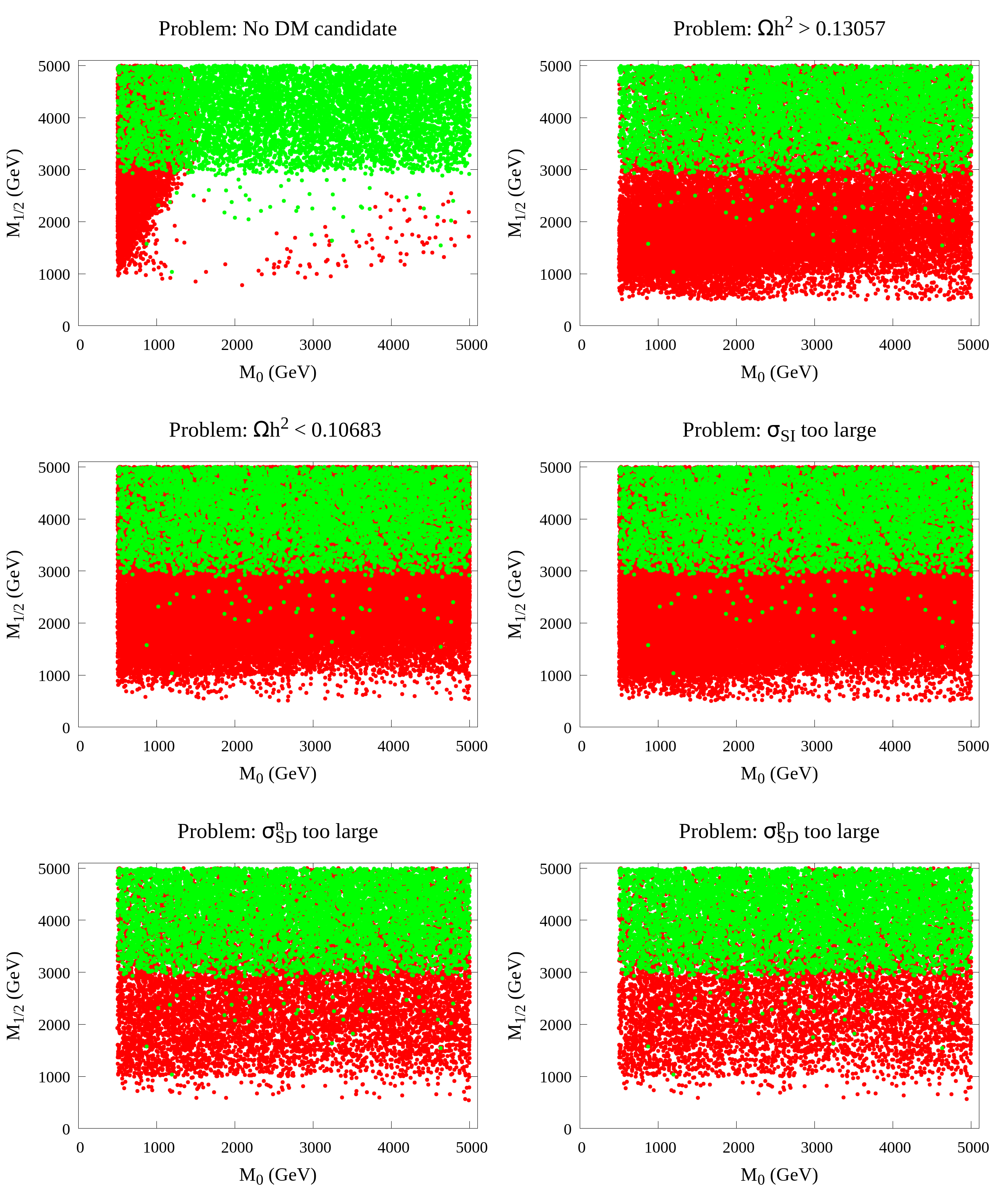}
 \caption{DM constraints in the $m_0$--$m_{1/2}$ plane. Red points are ruled out; green points are allowed.}
 \label{fig3}
\end{figure*}

Figures \ref{fig1}, \ref{fig2}, and \ref{fig3} show the results in the $(A_{\lambda}$-$A_{\kappa})$, $(\lambda$-$\kappa)$, and $(m_0$-$m_{1/2})$ planes, respectively. Each figure contains a number of sub-figures that represent the allowed points in green and the ruled-out points in red. Each sub-figure also represents one type of problem causing the red points to be ruled out.

In Figure~\ref{fig1} in the top left panel, the ruled-out points are associated with no DM candidate in the model. Basically, this reflects results where the LSP is a charged particle (hence, cannot be a DM candidate) or that the input parameters did not lead to successful solutions, causing the algorithm of MicrOMEGAs to stop. This issue becomes more relevant as both $A_{\lambda}$ and $A_{\kappa}$ increase, especially for $A_{\lambda} > -2000$ GeV. The top right panel shows the ruled-out points that predict too-large values of the relic density $\Omega h^2 \gtrsim 0.131$, whereas the middle left panel shows red points with $\Omega h^2 \lesssim 0.107$. Comparing the two plots indicates that in the scanned parameter space, most of the problematic points (passing all non-DM constraints still) are associated with too-small values for DM relic density\footnote{Such models might be accepted if DM is not provided by the LSP of the NMSSM. However, other DM constraints should be considered to agree with observations.}. As for cross-sections, the middle left panel presents models predicting a DM direct detection rate that is too large, that is, a larger than experimentally allowed spin-independent cross-section $\sigma_{SI}$, which resembles the density of ruled-out points due to $\Omega h^2$ being too small in the previously described plot. This shows that both of these limits play a major role in constraining the scNMSSM, and it warns against the common practice of imposing the upper limit on $\Omega h^2$ only, since even if a point with too-small relic density is accepted, such a point might be ruled out by direct detection constraints. Next, the ruled-out points due to spin-dependent cross-sections, $\sigma^n_{SD}$ and $\sigma^p_{SD}$ being larger than the experimental limits, are shown in the bottom left and bottom right panels, respectively. These constraints have less of an impact, but can eliminate parts of the model where both $A_{\lambda}$ and $A_{\kappa}$ are positive. In all the previous sub-figures, the DM constraints heavily affect regions where both trilinears are positive and large. 
Lastly, and focusing on the allowed points shown in green, the figure shows that when $A_{\lambda} < 0$, only negative $A_{\kappa}$ can lead to successful predictions. However, as $A_{\lambda}$ becomes positive, $A_{\kappa}$ can take positive values.

To facilitate comparison and acquire deeper insight, the same results are presented in the $\lambda$--$\kappa$ plane, as shown in Figure~\ref{fig2}. Some of the data passing all non-DM restrictions still failed to produce a successful DM candidate (top left panel). This occurred particularly in the region $0.4<\lambda<0.7$ and $0.3<\kappa<0.6$. Other points failed to comply with the upper limit on $\Omega h^2$ (top right panel), which occurred in all of the parameter space, but, as shown, regions where $\lambda \gtrsim 0.55$ with $\kappa \lesssim 0.2$ suffer from this problem as no allowed points can be found there (given our choices of inputs). The middle left panel shows the case where $\Omega h^2$ is too small. Comparing this result with that of too-large $\sigma_{SI}$, these two problems produce the main impact, as stated before. This impact is reflected in the $\lambda$--$\kappa$ plane for large values of $\lambda$ accompanied with small values of $\kappa$. In the same region, in both the bottom left and right panels, the limits of spin-dependent cross-sections affect the same bad region. 

%\afterpage{%
%\clearpage
%}

%\afterpage{%
%\clearpage
%}

%\afterpage{%
%\clearpage
%}

Figure~\ref{fig3} shows the results in the common $m_0$-$m_{1/2}$ plane, which in turn provides further understanding of how the DM constraints impact the parameter space of the model, and enables comparison with other low-scale SUSY models such as the MSSM. In the top left panel, the issue of not obtaining a successful DM candidate is concentrated in regions where $m_0 \lesssim 1200$ GeV. Most of the ruled-out points due to predicting large $\Omega h^2$ are associated with $m_{1/2} < 3000$ GeV (top right panel). The two main problems causing the ruling-out of the red points, that is, small $\Omega h^2$ and large $\sigma_{SI}$, are common in the whole parameter space (see middle panels), whereas the limits on the spin-dependent cross-sections are mainly present in the region where $m_{1/2} < 3000$ GeV. However, these are somewhat less relevant than the previous two constraints that occur in the same area. These results also show that the good points (in green) are concentrated in regions where $m_{1/2} > 3000$ GeV. Such a large value for $m_{1/2}$, which is needed to satisfy other non-DM limits, indicates a heavy mass spectrum, which is certainly the case in the studied parameter space.

\subsubsection{The Allowed Parameter Space in the $m_0$--$m_{1/2}$ Plane}

\begin{figure*} 
\centering
 \includegraphics[width=1\textwidth]{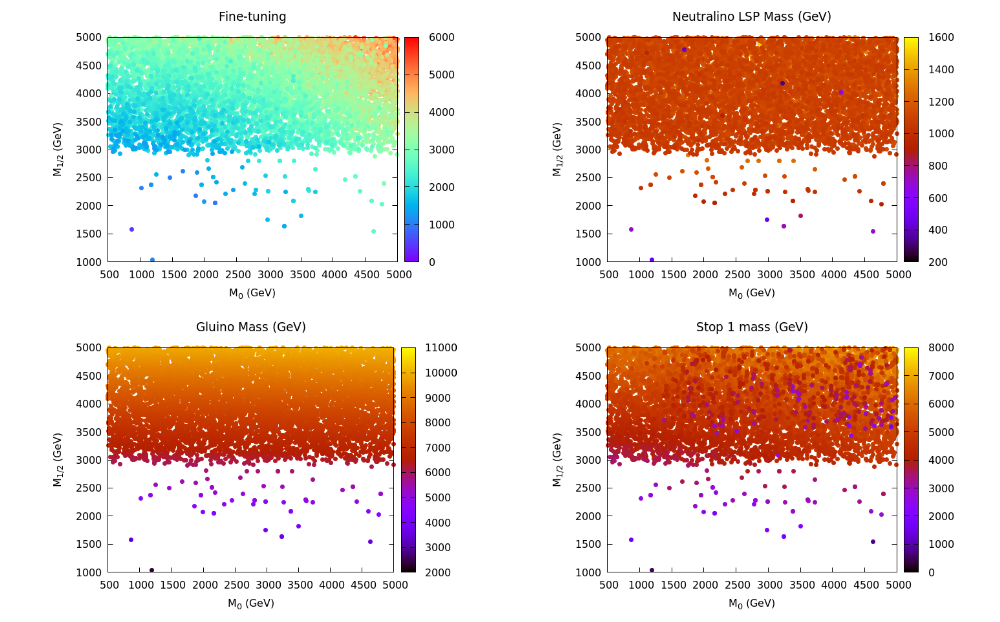}
 \caption{The allowed parameter space in the $m_0$--$m_{1/2}$ plane.}
 \label{figft}
\end{figure*}

\begin{figure*} 
\centering
 \includegraphics[width=1\textwidth]{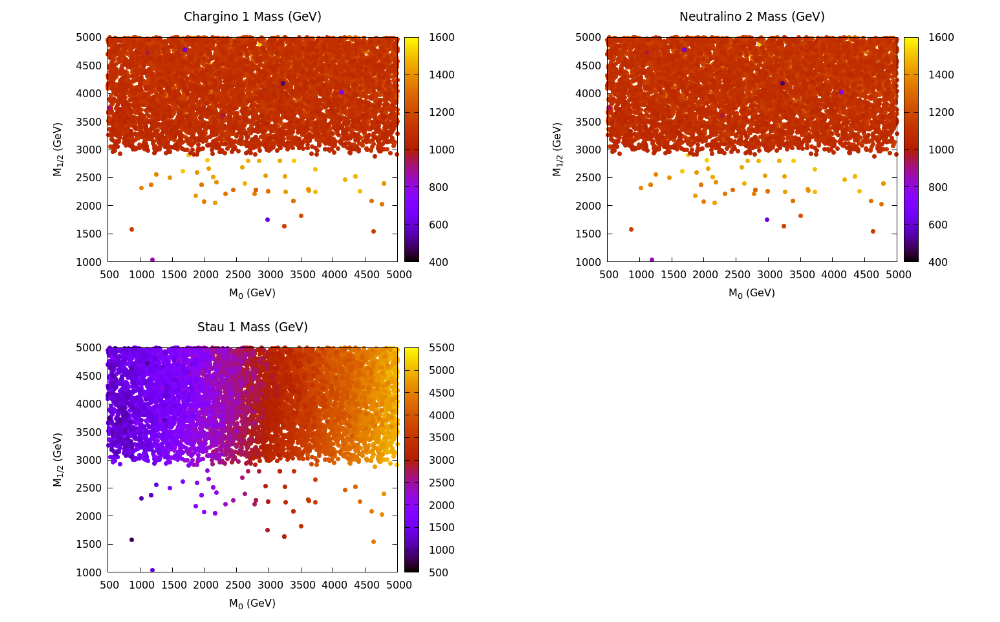}
 \caption{The allowed parameter space in the $m_0$--$m_{1/2}$ plane.}
 \label{figc}
\end{figure*}

 In the NMSSM, either $h_1$ or $h_2$ can be SM-like. However, in the parameter space we cover, the SM-like Higgs is $h_1$, the lightest CP-even Higgs boson. The mass varies between 122.1 and 128.1 GeV to account for theoretical uncertainties, as mentioned in the previous section.
To understand the allowed parameter space and enable comparison with the literature and future studies, Figures \ref{figft} and \ref{figc} present the range of the fine tuning in the model, as well as the mass range of the DM LSP particle ($m_{\tilde{\chi}_1^0}$), gluino ($m_{\tilde{g}}$), lightest stop ($m_{\tilde{t}_1}$), lightest chargino ($m_{\tilde{\chi}_1^\pm}$), the second-lightest neutralino ($m_{\tilde{\chi}_2^0}$), and the lightest stau ($m_{\tilde{\tau}_1}$). 

Figure~\ref{figft} in the top left panel shows that the fine-tuning ranges from $\Delta \sim$ 390 to 5000. As expected, regions where both $m_0$ and $m_{1/2}$ are as small as possible have lower fine tuning. The point with the smallest fine-tuning ($\Delta \sim386)$ in the scanned parameter space is associated with $m_0=870$ GeV, $m_{1/2}=1576$ GeV, and a bino-like DM candidate $m_{\tilde{\chi}_1^0} \sim693$ GeV. The details of the other inputs, the mass spectrum, the relevant parameters, and DM observables are presented in Tables \ref{table1} and \ref{table2}, where this point is denoted $\textbf{P1}$. 
The top right panel shows the mass of the DM LSP, which has a minimum mass of $m_{\tilde{\chi}_1^0} \sim313$ GeV and a maximum mass of 1526 GeV. The smallest DM candidate in the scanned parameter space is singlino-like. This point is denoted $\textbf{P4}$ and its details are presented in the previously mentioned tables.
The rest of the plots in Figures \ref{figft} and \ref{figc} show that the mass of the gluino, lightest stop, lightest chargino, sencond-lightest neutralino, and the lightest stau range within $[2326-10,050]$, $[489-7202]$, $[487-1528]$, $[491-1532]$, and $[787-5239]$ GeV, respectively. 
Finally, in the allowed parameter space, the lightest chargino is the next-to-lightest supersymmetric particle (NLSP), and, in general, the mass of the LSP is linearly correlated with the lightest chargino and the second-lightest neutralino.

\subsection{Properties of the Mass of the Neutralino DM}
Now that we have an understanding of how recent DM constraints affect the parameters of the model, it is crucial to analyze the DM particle in terms of its mass content. In the allowed parameter space, the LSP represents a valid DM candidate and has a mass range that is mostly concentrated between 1000 and 1500 GeV. Lighter and heavier values still exist in our results, but such cases are rare. 

\begin{figure*}[h!] 
\centering
 \includegraphics[width=0.85\textwidth]{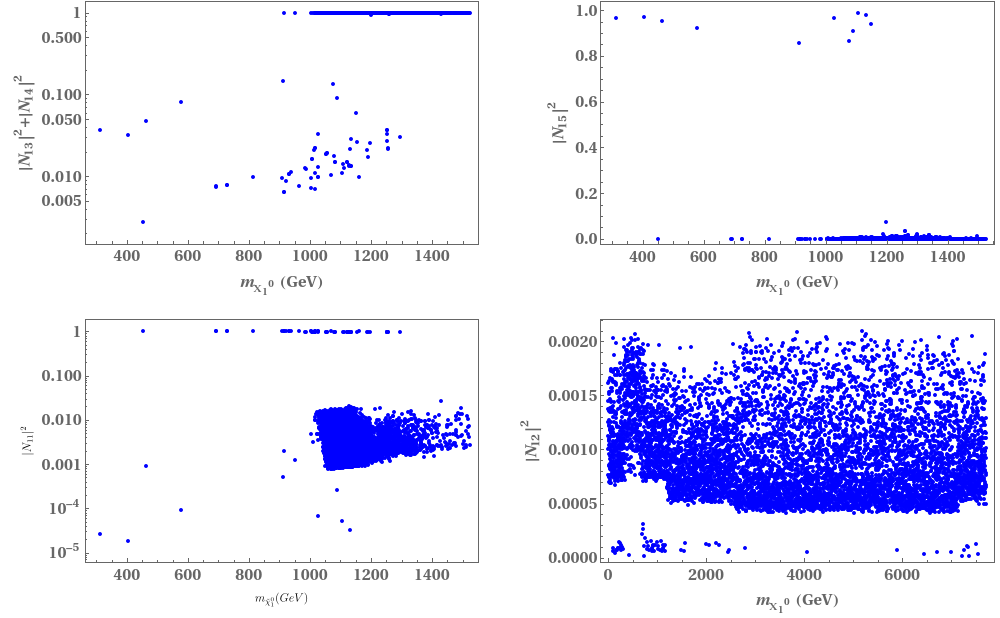}
 \caption{Mass components of $\tilde{\chi}_1^0$. }
 \label{fig4}
\end{figure*}

Figure~\ref{fig4} shows the LSP mass components as a function of LSP mass. The top left panel represents the higgsino component, where the points split into two regions, having large and small contributions to the LSP mass. The region with higher contribution occurs in the range of $916\leq m_{\tilde{\chi}_1^0}\leq 1526$ GeV with a maximum contribution of $|N_{13}|^2 + |N_{14}|^2 \sim0.99$, associated with $m_{\tilde{\chi}_1^0} \approx 1077$ GeV. This point in the parameter space is denoted $\textbf{P2}$ and is presented in Tables \ref{table1} and \ref{table2}. 

The singlino component as a function of LSP mass is shown in the top right panel. Again, the data points split into two regions with high and low singlino contribution to the LSP mass. The region with higher contribution occurs between $313\leq m_{\tilde{\chi}_1^0}\leq 1149$ GeV. A representative point ($\textbf{P4}$) where the singlino contribution is $|N_{15}|^2 \approx 0.99$ and $m_{\tilde{\chi}_1^0} \approx 313$ GeV is presented in Tables \ref{table1} and \ref{table2}. It is possible to find cases where the DM particle is a singlino--higgsino mixture. For instance, in the allowed parameter space, we find a point where $|N_{15}|^2 \approx 0.85$ and $|N_{13}|^2 + |N_{14}|^2 \approx 0.14$. This point is denoted $\textbf{P3}$ in the previously mentioned tables shown the Appendix.

The bottom right panel of Figure~\ref{fig4} shows that the $W$ino contribution to the LSP is always small in the explored parameter space. This region is associated with $313\leq m_{\tilde{\chi}_1^0}\leq 1525$ GeV, where the low and large mass values are associated with $|N_{12}|^2 \lesssim 0.0021$. 

The bino components are given in the bottom left panel of Figure ~\ref{fig4}, where the data points split into two regions with higher and lower contributions to the LSP mass. The region with large bino contribution occurs for $451\leq m_{\tilde{\chi}_1^0}\leq 1294$ GeV, where the value of $|N_{11}|^2$ can approach $0.99$. A representative point of this case is provided in the Appendix. However, in most of the parameter space, the value of the bino component is below 0.05.

Finally, the results above show that in the parameter space under consideration, the LSP is mostly higgsino for the majority of data points. Cases where it is mostly bino, singlino, and singlino--higgsino also exist in our data. The $W$ino component is always negligible. To illustrate the physical properties of such cases, we present four representative points: $\textbf{P1}$ (bino DM), $\textbf{P2}$ (higgsino DM), $\textbf{P3}$ (singlino-higgsino DM), and $\textbf{P4}$ (singlino DM) in Tables \ref{table1} and \ref{table2}.

% \afterpage{%
% \clearpage
% }

\subsection{Contributions to Dark Matter Relic Density}

Several annihilation and co-annihilation processes contribute to dark matter relic density. Here, we present the results for the processes that can contribute more than 1\%. In the relevant plots, all of the points with a zero y-axis correspond to data points that have a contribution of less than 1\%. The exact values were not included in the output routine since these are very small (and hence there were set to zero in the plotting command), but those values are considered in MicrOMEGAs and NMSSMTools for the calculation of the thermally averaged cross-section and the corresponding relic density. 

\begin{figure*}
\centering
 \includegraphics[width=0.85\textwidth]{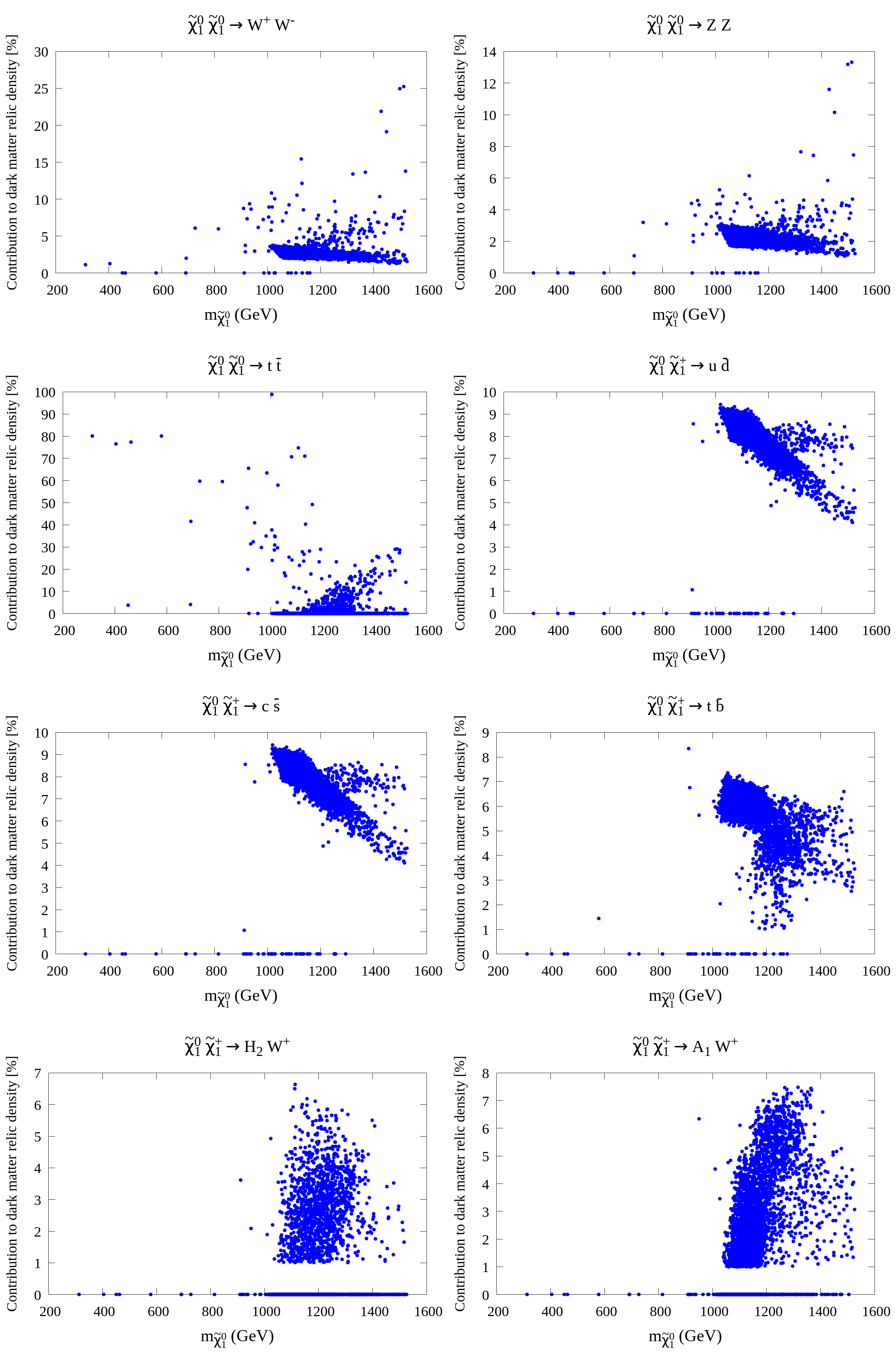}
 \caption{Contributions of annihilation and co-annihilation processes to dark matter relic density.}
 \label{fig5}
\end{figure*}

Figure~\ref{fig5} illustrates the percent contribution to DM relic density for the annihilation and co-annihilation modes. The annihilation mode $\tilde{\chi}_1^0\tilde{\chi}_1^0\rightarrow W^+W^-$ is given in the top left panel, where the data are spread between $ m_{\tilde {\chi}_1^0}=315$ GeV at the lower end, corresponding to a contribution of about 1\$, and $ m_{\tilde {\chi}_1^0}=1526$ GeV at the higher end, corresponding to a contribution of about 1.5\%. Most of the data points cover a mass range of $m_{\tilde {\chi}_1^0}=1018-1350$ GeV with contribution of 1.72--3.66\%. The largest contribution from this process is about 25\%, corresponding to $m_{\tilde {\chi}_1^0}=1516$ GeV, and the smallest contribution is about 1\%, corresponding to $m_{\tilde {\chi}_1^0}=315$ GeV.

The annihilation channel $\tilde{\chi}_1^0\tilde{\chi}_1^0\rightarrow ZZ$ is shown in the top right panel. In this case, the data points spread from a lower value of $ m_{\tilde {\chi}_1^0}=693$ GeV, corresponding to a contribution of $1.1\%$, to a higher value of $ m_{\tilde {\chi}_1^0}=1526$ GeV, corresponding to the contribution 1.24\%. The maximum contribution is 13.2\%, which corresponds to $ m_{\tilde {\chi}_1^0}=1514$ GeV, while the minimum contribution is 1.1\%, corresponding to $ m_{\tilde {\chi}_1^0}=1487$ GeV. The majority of data points lie in the range $m_{\tilde {\chi}_1^0}=1028-1347$ GeV with contributions between 2.95\% and 1.57\%. 

The left panel in the second row shows the contribution of the annihilation mode $\tilde{\chi}_1^0\tilde{\chi}_1^0\rightarrow t\bar{t}$ to DM relic density. In this case, the lower end of the mass starts at $ m_{\tilde {\chi}_1^0}=314$ GeV, corresponding to a contribution of 80\%, while the higher mass value is $ m_{\tilde {\chi}_1^0}=1522$ GeV, corresponding to a contribution 14\%. The maximum contribution to the relic density is 99\%, corresponding to an LSP mass of $ m_{\tilde {\chi}_1^0}=1007$ GeV. Most of the data points occur in the range $m_{\tilde {\chi}_1^0}=1161-1290$ GeV, with contributions in the range of 5.75--0.69\%, where 0.69\% represents the minimum contribution.

The co-annihilation modes $\tilde{\chi}^0_1 \tilde{\chi}^+_1 \rightarrow u\bar{d},c\bar{s}$ are illustrated in the right panel of the second row and the left panel of the third row, respectively. Here, the data points are spread between a lower value of $m_{\tilde {\chi}_1^0}=913$ GeV, corresponding to a contribution 1.1\%, and a higher value of $m_{\tilde {\chi}_1^0}=1528$ GeV, corresponding to a contribution 5\%. The maximum contribution is 9.4\%, which corresponds to $m_{\tilde {\chi}_1^0}=1019$ GeV; the minimum contribution is 1.1\%, corresponding to $m_{\tilde {\chi}_1^0}=913$ GeV. Most of the allowed points are located in the range of $m_{\tilde {\chi}_1^0}=1029-1330$ GeV, with a contribution range of 7--9\%.

The co-annihilation mode $\tilde{\chi}^0_1 \tilde{\chi}^+_1 \rightarrow t\bar{b}$ is given in the right panel of the third row. The allowed points are spread between a lower $m_{\tilde {\chi}_1^0}=580$ GeV, corresponding to a contribution 1.45\%, and a higher value of $m_{\tilde {\chi}_1^0}= 1526$ GeV, corresponding to a contribution of 3.46\%. The maximum contribution is 8.36\%, which corresponds to $m_{\tilde {\chi}_1^0}=914$ GeV, whereas the minimum contribution is 1.01\%, corresponding to $m_{\tilde {\chi}_1^0}=1195$ GeV. Most of the data points are located at $m_{\tilde {\chi}_1^0}=1023-1400$ GeV, with a contribution range of 3.41--7.31\%.	%In ranges, low value is usually provided first.

The bottom left panel shows the co-annihilation channel $\tilde{\chi}^0_1 \tilde{\chi}^+_1 \rightarrow H_2W^+$. In this case, the data points are spread between $m_{\tilde {\chi}_1^0}=914$ GeV, corresponding to a contribution of 3.62\%, and $m_{\tilde {\chi}_1^0}=1514$ GeV, corresponding to a contribution of 1.66\%. The maximum contribution is 6.63\%, which corresponds to $m_{\tilde {\chi}_1^0}=1114$ GeV. The majority of data points are located at $m_{\tilde {\chi}_1^0}=1069-1324$ GeV, with a contribution range of 0.998--4.59\%.

The last co-annihilation mode considered here is $\tilde{\chi}^0_1 \tilde{\chi}^+_1 \rightarrow A_1W^+$, which is shown in the bottom right panel. The data points are spread between $m_{\tilde {\chi}_1^0}=952$ GeV, corresponding to a contribution of 6.35\%, and $m_{\tilde {\chi}_1^0}=1526$ GeV, corresponding to a contribution of 3.1\%. The maximum contribution is 7.1\%, associated with $m_{\tilde {\chi}_1^0}=1317$ GeV. Most data points are located at $m_{\tilde {\chi}_1^0}=1043-1338$ GeV, with a contribution range of 0.97--6.82\%.

%\afterpage{%
%\clearpage
%}
\subsection{$\tilde{\chi}_2^0$ and $\tilde{\chi}_1^{\pm}$ Decays to the LSP}

Several decay channels produce an LSP (along with other particles); however, the decays of the chargino and the second-lightest neutralino are of special importance.

\begin{figure*}[h!]
\centering
 \includegraphics[width=0.85\textwidth]{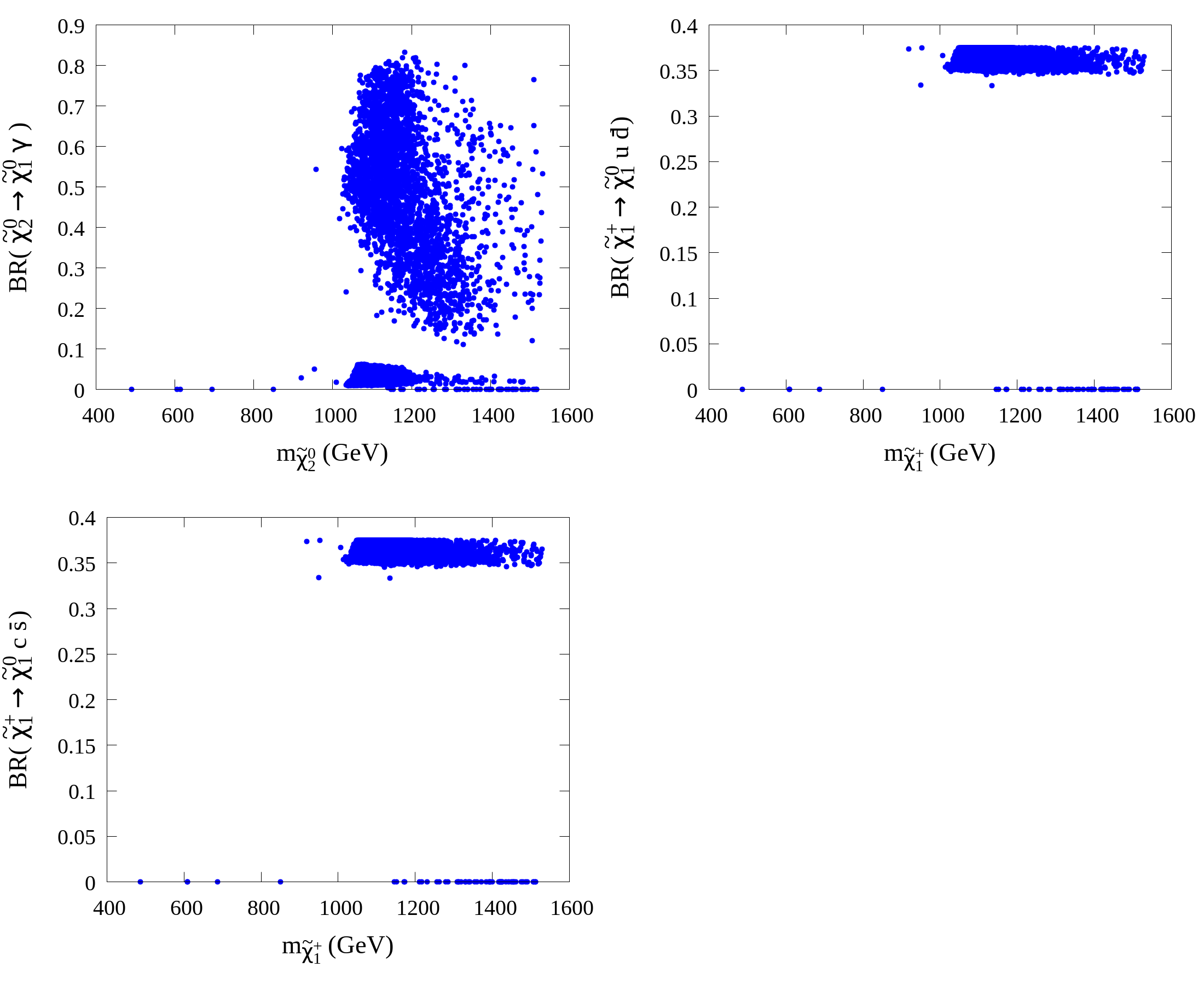}
 \caption{Branching ratios of $\tilde{\chi}_2^0$ and $\tilde{\chi}_1^+$ to $\tilde{\chi}_1^0$ and other particles, where only decays with sizeable $Br$ throughout the parameter space are shown.}
 \label{fig6}
\end{figure*}

Figure~\ref{fig6} shows that for $\tilde{\chi}_2^0$ decays, the radiative decay $\tilde{\chi}_2^0\rightarrow \tilde{\chi}_1^0 \gamma$ can be significant in regions of the parameter space, where the branching ratio can reach up to $\sim0.8$ for a nearly degenerate mass of $\sim1200$ GeV for both $\tilde{\chi}_2^0$ and $\tilde{\chi}_1^0$. An important type of decay channel for $\tilde{\chi}_1^+$ is $\tilde{\chi}_1^+\rightarrow \tilde{\chi}_1^0 u\bar{d}/c\bar{s}$, where the branching ratio lies in the range of $0.35 \leq BR(\tilde{\chi}_1^+\rightarrow \tilde{\chi}_1^0 u\bar{d})\leq 0.38$ for almost all of the allowed parameter space.

Other decay modes of both considered sparticles into the LSP can be important as well. For instance, the decay $\tilde{\chi}_2^0\rightarrow \tilde{\chi}_1^0 Z$ can be important, but in very rare cases where the branching ratio is equal to 1 (where the difference in mass between both neutralinos is around 100 GeV), which does not represent the region where the majority of allowed points lie. Additionally, decays of $\tilde{\chi}_2^0$ to a pair of SM $q\bar{q}$ and LSP can be significant, whereas weakly mediated decays of $\tilde{\chi}_1^+$ to an LSP and a pair of $f\bar{f}$ (where such a pair can consist of a lepton/antilepton and an antineutrino/neutrino) can also be important with a branching ratio above 0.3. For a mass range of $1154\leq m_{\tilde{\chi}_2^0}\leq 1517$ GeV, the decay channel $\tilde{\chi}_2^0\rightarrow \tilde{\chi}_1^0 H_1$ has a branching ratio that is close to unity $0.94\leq BR(\tilde{\chi}_2^0\rightarrow \tilde{\chi}_1^0 H_1)\leq 1$ for a small number of successful points (where the mass difference between the two neutralinos is around a minimum of 200 GeV). The results of the branching ratios for $\tilde{\chi}_1^+\rightarrow \tilde{\chi}_1^0 W^+$ indicate that the branching ratio is equal to 1, mostly in the range of $m_{\tilde{\chi}_1^+}=1151-1513$ GeV for a limited number of successful points (where the mass difference between the chargino and the neutralino is about a minimum of 84 GeV). The lower value $BR(\tilde{\chi}_1^+\rightarrow \tilde{\chi}_1^0 W^+)= 0.033$ corresponds to $m_{\tilde{\chi}_1^+}=849$ GeV. 

Finally, in the scanned parameter space, some of the remaining decay channels of $\tilde{\chi}_2^0$ and $\tilde{\chi}_1^+$ into $\tilde{\chi}_1^0$ have a range of branching ratios that reaches a maximum value of $Br(\tilde{\chi}_1^+ \rightarrow \tilde{\chi}_1^0 e^+ \nu_e) \lesssim 0.13$ , $Br(\tilde{\chi}_1^+ \rightarrow \tilde{\chi}_1^0 \mu^+ \nu_{\mu})\lesssim 0.12$, $Br(\tilde{\chi}_1^+ \rightarrow \tilde{\chi}_1^0 \tau^+ \nu_{\tau})\lesssim 0.11$, and $Br(\tilde{\chi}_2^0 \rightarrow \tilde{\chi}_1^0 c \bar{c}) < 0.14$, $Br(\tilde{\chi}_2^0 \rightarrow \tilde{\chi}_1^0 d \bar{d})< 0.18$, $Br(\tilde{\chi}_2^0 \rightarrow \tilde{\chi}_1^0 \nu_{e/\mu/\tau} \bar{\nu}_{e/\mu/\tau}) \lesssim 0.08$, $Br(\tilde{\chi}_2^0 \rightarrow \tilde{\chi}_1^0 s \bar{s})< 0.18$, and $Br(\tilde{\chi}_2^0 \rightarrow \tilde{\chi}_1^0 u \bar{u})< 0.14$. Other channels are quite rare (i.e., very small branching ratio) or no decays were found in the scanned parameter space. These include $\tilde{\chi}_2^0 \rightarrow \tilde{\chi}_1^0 e^+ e^-$, $\tilde{\chi}_2^0 \rightarrow \tilde{\chi}_1^0 \mu^+ \mu^-$, $\tilde{\chi}_2^0 \rightarrow \tilde{\chi}_1^0 \tau^+ \tau^-$, $\tilde{\chi}_2^0 \rightarrow \tilde{\chi}_1^0 b \bar{b}$, and $\tilde{\chi}_2^0 \rightarrow \tilde{\chi}_1^0 A_1/A_2$, where the branching ratio reaches maximum values well below 0.1, and $\tilde{\chi}_2^0 \rightarrow \tilde{\chi}_1^0 t \bar{t}$, $\tilde{\chi}_2^0 \rightarrow \tilde{\chi}_1^0 H_2/H_3$, and $\tilde{\chi}_1^+ \rightarrow \tilde{\chi}_1^0 t \bar{b}$, which are not found in the scanned parameter space.

% \afterpage{%
% \clearpage
% }

\section{Discussion and conclusions} \label{sec:discon}
The results in the previous section showed that the NMSSM with semi-constrained GUT boundary conditions can explain current experimental observations including DM. However, the mass spectrum of the new particles, especially $m_{\tilde{g}}$ and $m_{\tilde{t}_1}$, can be heavy. As far as the input parameters are concerned, large values of $m_0$ and $m_{1/2}$ are required. As such inputs increase, the whole SUSY spectrum increases, including the LSP dark matter. There are several reasons for this situation, the most crucial of which is satisfying the bounds from direct searchers for SUSY particles. We can still look for regions where the parameters of the model conspire to produce a lighter spectrum (e.g., regions with specific relations between the gluino mass $m_{\tilde{g}}$ or the stop mass $m_{\tilde{t}_1}$ and $m_{\tilde{\chi}_1^0}$), but we did not consider these cases in this paper. 

The effects of DM constraints on other NMSSM-specific parameters were considered. These parameters include $\lambda$ and $\kappa$, where we found that they can assume values between 0.1 and below 0.7; however, the majority of allowed points reside in regions where $\lambda \lesssim 0.6$ and $\kappa \gtrsim 0.3$. The other NMSSM-specific parameters, $A_{\lambda}$ and $A_{\kappa}$, for most of the allowed parameter space are found to lay mostly in regions where $A_{\kappa}$ is negative. 

In the scanned parameter space, we observed that the DM particle (the neutralino LSP) is mostly higgsino for a range of neutralino mass between 1000 and 1500 GeV (e.g., $\textbf{P2}$ ,where $|N_{13}|^2 + |N_{14}|^2 \approx 0.99$ and $m_{\tilde{\chi}_1^0} \approx 1077$ GeV). The singlino component was found to be quite small in the parameter space. However, this does not rule out the option of finding regions where the singlino component could be large. We noted that some points in the scanned parameter space achieved such large values (e.g., $\textbf{P4}$, where $|N_{15}|^2 \approx 0.99$ and $m_{\tilde{\chi}_1^0} \approx 313$ GeV). Further exploration of this case requires a dedicated and separate study, as this was beyond the scope of this paper. Furthermore, we found that some points have a DM candidate that is a singlino--higgsino mixture (e.g., $\textbf{P3}$, where $|N_{15}|^2 \approx 0.85$, $|N_{13}|^2 + |N_{14}|^2 \approx 0.14$, and $m_{\tilde{\chi}_1^0} \approx 911$ GeV). The bino component was found to be somewhat small in most of the allowed parameter space. Again, we found that some points have large bino components (e.g., $\textbf{P1}$, where $|N_{11}|^2 \approx 0.99$ and $m_{\tilde{\chi}_1^0} \approx 693$ GeV), and investigating such regions is an interesting aspect for future research. The last component, the wino, was found to be negligible throughout the allowed parameter space. 
We observed that in most of the allowed parameter space, the annihilation of the LSP, as a mechanism for generating the required DM relic density, is not sufficient in the model. To achieve the observed value, co-annihilation must be considered. We found that co-annihilation processes with the chargino $\tilde{\chi}_1^+$ constitute a major contributor to the dark matter relic density given the specific allowed parameter space. Specifically, co-annihilation processes leading to $u \bar{d}/c\bar{s}/t\bar{b}$, $H_2 W^+$, and $A_1 W^+$ were found to provide most of the contributions. 

Compared to previous literature, we note that the requirement that $(g-2)_{\mu}$ is explained by the model in \cite{Wang:2018vxp, Wang:2020tap} has restricted the maximum values of $m_0$ and $\mu_{\text{eff}}$ to 500 GeV and 200 GeV. In that region of the parameter space it was found in these two references that $m_{h_2}$ is the SM-like Higgs, whereas in our work we have found that in the wider parameter space that we have covered, it is $m_{h_1}$ that becomes the SM-like Higgs. Moreover, in the limited range of parameter space that was covered in \cite{Wang:2018vxp, Wang:2020tap}, only singlino-dominated or higgsino-dominated DM were found, some of which may well have relic density below the experimental limit as only the upper limit on DM relic density was imposed by the authors, whereas in our scanned parameter space we have found singlino-like, higgsino-like, higgsino-singlino-like, and bino-like DM cases as mentioned above, all of which can fully account for the DM relic density. As a final note, in both references, the mass of the DM particle is always smaller than $O(100)$ GeV, whereas in our results we have covered a wider range of predictions for the DM mass. Therefore, our results provide a wider view of the case of DM in the scNMSSM.

Focusing on our results for the decays of heavier neutralinos and charginos to the LSP, we found that the decays of $\tilde{\chi}_2^0$ and $\tilde{\chi}_1^+$ are crucial as they can be close to the mass of $\tilde{\chi}_1^0$. We observed that the radiative decay of $\tilde{\chi}_2^0$ can be significant, accounting for about 80\% of its decays, especially in regions where it has a nearly degenerate mass with $\tilde{\chi}_1^0$, which is the region where most allowed points were found and the common mass between the two was about 1200 GeV. Other channels can be significant, albeit very rarely, as we discussed in the results.

In summary, the NMSSM with GUT boundary conditions remains an interesting model for new physics. We showed that it can account for current observations, even though the predicted spectrum appears to be heavy. In the future, a dedicated study of regions where the DM particle is mostly singlino could be conducted to analyze the viability of this case given the most recent stringent constraints.

\section*{Acknowledgements}
The authors would like to thank the Deanship of Scientific Research in King Saud University for funding and supporting this research through the initiative of DSR Graduate Students Research Support (GSR). Maien Binjonaid would like to thank Constantinos Pallis for discussions at an early stage of this work.

\section*{Appendix} \label{app}
Here, we present four points in the parameter space where the DM candidate is bino-like ($\textbf{P1}$), which is the point with lowest fine-tuning in the scanned parameter space, higgsino-like ($\textbf{P2}$), singlino--higgsino-like ($\textbf{P3}$), and singlino-like ($\textbf{P4}$), which is the point with the smallest mass for the DM candidate.

\begin{table} 
	\centering
\begin{tabular}{|c|c|c|c|c|}
\hline 
 & P1 & P2 & P3 & P4 \tabularnewline
\hline 
LSP Type & Bino & Higgsino & Singlino--Higgsino & Singlino \tabularnewline
\hline 
\hline
$m_0$ (GeV) & 870 & 2028 & 2285 & 3227 \tabularnewline
\hline 
$m_{1/2}$ (GeV) & 1576 & 4992 & 3600 & 4185\tabularnewline
\hline 
$A_0$ (GeV) & 25 & --2700 & --1336 & --2077\tabularnewline
\hline 
$A_{\lambda}$ (GeV) & --150 & --2428 & 0 & 0\tabularnewline
\hline 
$A_{\kappa}$ (GeV) & --710 & --2700 & 0 & 0\tabularnewline
\hline 
$\lambda$ & 0.51 & 0.13 & 0.50 & 0.48\tabularnewline
\hline 
$\kappa$ & 0.46 & 0.58 & 0.24 & 0.15\tabularnewline
\hline 
$\tan{\beta}$ & 1.43 & 9.86 & 1.38 & 2.17\tabularnewline
\hline 
$\mu_{\text{eff}}$ (GeV) & 1190 & 1051 & 940 & 474\tabularnewline
\hline
\hline
$m_{\tilde{\chi}_1^0}$ (GeV) & 693 & 1077 & 911 & 313\tabularnewline
\hline 
$N_{11}$ & 0.99 & 0.03 & 0.02 & 0.0051\tabularnewline
\hline 
$N_{12}$ & --0.011 & --0.02 & --0.01 & --0.0047\tabularnewline
\hline 
$N_{13}$ & 0.064 & 0.71 & 0.23 & 0.05\tabularnewline
\hline 
$N_{14}$ & -0.058 & --0.71 & --0.29 & --0.18\tabularnewline
\hline 
$N_{15}$ & 0.0012 & 0.002 & 0.93 & 0.98\tabularnewline
\hline 
$m_{\tilde{\chi}_2^0}$ (GeV) & 1153 & 1080 & 952 & 490\tabularnewline
\hline 
$m_{\tilde{\chi}_3^0}$ (GeV) & 1193 & 2274 & 956 & 496\tabularnewline
\hline 
$m_{\tilde{\chi}_4^0}$ (GeV) & 1349 & 4143 & 1642 & 1895\tabularnewline
\hline 
$m_{\tilde{\chi}_5^0}$ (GeV) & 2153 & 9687 & 3004 & 3480\tabularnewline
\hline 
$m_{\tilde{\chi}_1^{\pm}}$ (GeV) & 1150 & 1079 & 950 & 487\tabularnewline
\hline 
$m_{\tilde{\chi}_2^{\pm}}$ (GeV) & 1347 & 4143 & 3004 & 3480\tabularnewline
\hline 
$m_{h_1}$ (GeV) & 122.5 & 127.4 & 124.2 & 123.6\tabularnewline
\hline 
$m_{H_2}$ (GeV) & 1910 & 5990 & 790 & 292\tabularnewline
\hline 
$S_{23}$ & 0.01 & 0.004 & 0.99 & 0.99\tabularnewline
\hline 
$m_{H_3}$ (GeV) & 2101 & 9199 & 1721 & 1109\tabularnewline
\hline 
$m_{A_1}$ (GeV) & 713 & 3279 & 780 & 210\tabularnewline
\hline 
$P_{13}$ & 0.99 & 0.99 & 0.99 & 0.99\tabularnewline
\hline 
$m_{A_2}$ (GeV) & 1913 & 5990 & 1720 & 1108\tabularnewline
\hline 
$m_{H^\pm}$ (GeV) & 1903 & 5994 & 1704 & 1104\tabularnewline
\hline
$m_{\tilde{g}}$ (GeV) & 3378 & 9888 & 7336 & 8463\tabularnewline
\hline 
$m_{\tilde{t}_1}$ (GeV) & 1510 & 6210 & 2903 & 4870\tabularnewline
\hline 
$m_{\tilde{\tau}_1}$ (GeV) & 776 & 2495 & 2052 & 3126\tabularnewline
\hline 
$m_{\tilde{e}_R}$ (GeV) & 776 & 2564 & 2053 & 3127\tabularnewline
\hline 
$\Delta_{M_Z}^{\text{GUT}}$ & 383 & 3740 & 2100 & 3737\tabularnewline
\hline 
$\Omega h^2$ & 0.107 & 0.107 & 0.112 & 0.126\tabularnewline
\hline 
$\sigma_p(SI)$ (pb) & $5.977\times 10^{-10}$ & $1.836\times 10^{-10}$ & $1.492\times 10^{-10}$ & $6.226\times 10^{-11}$\tabularnewline
\hline 
$\sigma_p(SD)$ (pb) & $1.064 \times 10^{-8}$ & $1.140\times 10^{-7}$ & $4.282\times 10^{-5}$ & $4.277\times 10^{-5}$\tabularnewline
\hline 
$\sigma_n(SD)$ (pb) & $1.302 \times 10^{-8}$ & $8.716\times 10^{-8}$ & $3.274\times 10^{-5}$ & $3.271\times 10^{-5}$\tabularnewline
\hline 
\end{tabular}

\caption{All physical masses and mass parameters are in GeV and all cross-sections are in pb.}
\label{table1}
\end{table}

\begin{table} 
\centering
\begin{tabular}{|l|c|l|c|}
\hline 
\multicolumn{2}{|c|}{P1} & \multicolumn{2}{c|}{P2}\tabularnewline
\hline 
\multicolumn{2}{|c|}{Bino} & \multicolumn{2}{c|}{Higgsino}\tabularnewline
\hline 
Channel & Contribution (\%) & Channel & Contribution (\%)\tabularnewline
\hline 
$\tilde{\chi}_1^0 \tilde{\chi}_1^0\rightarrow t\bar{t}$ & 40 & $\tilde{\chi}_1^+ \tilde{\chi}_1^0 \rightarrow u \bar{d}/\bar{s} c$ & 8 \tabularnewline
\hline 
$\tilde{\chi}_1^0 \tilde{\chi}_1^0\rightarrow \tau \bar{\tau}/ \mu \bar{\mu} / e \bar{e}$ & 10 & $\tilde{\chi}_1^+ \tilde{\chi}_1^0 \rightarrow t \bar{b}$ & 7 \tabularnewline
\hline 
$\tilde{\chi}_1^0 \tilde{\chi}_1^0\rightarrow h_1 A_1$ & 9.2 & $\tilde{\chi}_1^+ \tilde{\chi}_2^0 \rightarrow u \bar{d} / \bar{s} c$ & 6 \tabularnewline
\hline 
$\tilde{\chi}_1^0 \tilde{\tau}_1 \rightarrow \gamma \tau$ & 4 & $\tilde{\chi}_1^+ \tilde{\chi}_2^0 \rightarrow t \bar{b}$ & 5.1 \tabularnewline
\hline 
$\tilde{\chi}_1^0 \tilde{e}_R \rightarrow \gamma e / \gamma \mu$ & 3.5 & $\tilde{\chi}_1^+ \tilde{\chi}_1^- \rightarrow W^+ W^-$ & 3 \tabularnewline
\hline 
$\tilde{\chi}_1^0 \tilde{\chi}_1^0\rightarrow W^+ W^-$ & 2 & $\tilde{\chi}_1^+ \tilde{\chi}_1^0 \rightarrow \nu_e \bar{e}/\nu_{\mu} \bar{\mu}/\nu_{\tau} \bar{\tau}$ & 2.6 \tabularnewline
\hline 
$\tilde{\chi}_1^0 \tilde{\chi}_1^0\rightarrow h_1 h_1$ & 1.2 & $\tilde{\chi}_1^+ \tilde{\chi}_1^- \rightarrow u \bar{u}/ c \bar{c}/ (t \bar{tt})$ &2.3 (2.1) \tabularnewline
\hline 
$\tilde{\chi}_1^0 \tilde{\tau}_1 \rightarrow Z \tau$ & 1 & $\tilde{\chi}_1^0 \tilde{\chi}_2^0 \rightarrow d \bar{d}/ s \bar{s} / b \bar{b}$ & 2.1 \tabularnewline
\hline 
& & $\tilde{\chi}_1^0 \tilde{\chi}_1^0\rightarrow W^+ W^-$ & 2 \tabularnewline
\hline 
& & $\tilde{\chi}_1^+ \tilde{\chi}_2^0\rightarrow \nu_e \bar{e}/ \nu_{\mu} \bar{\mu}/ \nu_{\tau} \bar{\tau}$ & 2 \tabularnewline
\hline 
\hline
\multicolumn{2}{|c|}{P3} & \multicolumn{2}{c|}{P4}\tabularnewline
\hline 
\multicolumn{2}{|c|}{Singlino--Higgsino} & \multicolumn{2}{c|}{Singlino}\tabularnewline
\hline
Channel & Contribution (\%) & Channel & Contribution (\%) \tabularnewline
\hline 
$\tilde{\chi}_1^0\tilde{\chi}_1^0 \rightarrow t \bar{t}$ & 23 & $\tilde{\chi}_1^0\tilde{\chi}_1^0 \rightarrow t \bar{t}$ & 78 \tabularnewline
\hline 
$\tilde{\chi}_1^+ \tilde{\chi}_1^0 \rightarrow t \bar{b}$ & 8 & $\tilde{\chi}_1^0\tilde{\chi}_1^0 \rightarrow h_2 A_1$ & 14 \tabularnewline
\hline 
$\tilde{\chi}_1^0\tilde{\chi}_1^0 \rightarrow Z h_2$ & 7.1 & $\tilde{\chi}_1^0\tilde{\chi}_1^0 \rightarrow h_1 A_1$ & 3.2 \tabularnewline
\hline 
$\tilde{\chi}_1^0\tilde{\chi}_1^0 \rightarrow h_2 A_1$ & 7 & $\tilde{\chi}_1^0\tilde{\chi}_1^0 \rightarrow W^+ W^-$ &1.2 \tabularnewline
\hline 
$\tilde{\chi}_1^0 \tilde{\chi}_1^0 \rightarrow W^\pm H^\pm$ & 4 & $\tilde{\chi}_1^0 \tilde{\chi}_1^0 \rightarrow Z h_2$ & 1 \tabularnewline
\hline 
$\tilde{\chi}_1^+ \tilde{\chi}_1^0 \rightarrow W^+ h_2$ & 3.3 & & \tabularnewline
\hline 
$\tilde{\chi}_1^0 \tilde{\chi}_3^0 \rightarrow t \bar{t}$ & 3 & & \tabularnewline
\hline 
$\tilde{\chi}_1^0 \tilde{\chi}_2^0 \rightarrow t \bar{t}$ & 2.2 & & \tabularnewline
\hline 
$\tilde{\chi}_1^+ \tilde{\chi}_1^0 \rightarrow Z W^+$ & 2 & & \tabularnewline
\hline 
$\tilde{\chi}_1^0 \tilde{\chi}_2^0 \rightarrow W^+ W^-$ & 1.4 & & \tabularnewline
\hline 
\end{tabular}
\caption{Percent contributions to the dark matter relic density for the four representative points.}
\label{table2}
\end{table}

\clearpage

\bibliography{references}
\bibliographystyle{unsrt}

\end{document}